\documentclass[twocolumn,twoside,slac_two]{revtex4}
\usepackage{graphicx}
\usepackage{fancyhdr}
\usepackage{threeparttable}
\usepackage{natbib}
\usepackage{amssymb}
\usepackage{epstopdf}
\begin{document}

\title{Optical Spectrophotometric Monitoring of Fermi/LAT Bright Sources}

\author{V. Pati\~no-\'Alvarez$^1$, V. Chavushyan$^1$, J. Le\'on-Tavares$^2$, J. R. Vald\'es$^1$, A. Carrami\~nana$^1$, L. Carrasco$^1$, J. Torrealba$^1$}
\affiliation{$^1$Instituto Nacional de Astrof\'isica, \'Optica y Electr\'onica, Luis Enrique Erro 1, Tonantzintla, Puebla, 72840, M\'exico\\ $^2$Finnish Centre for Astronomy with ESO (FINCA), University of Turku, V\"ais\"al\"antie 20, Piikki\"o, Turku, FI-21500, Finland} 

\begin{abstract}
\begin{small}
We describe an ongoing optical spectrophotometric monitoring program of a sample of Fermi/LAT bright sources showing prominent and variable $\gamma$-ray emission, with the 2.1m telescope at Observatorio Astrof\'isico Guillermo Haro (OAGH) located in Cananea, Sonora, M\'exico. Our sample contains 11 flat spectrum radio quasars (FSRQ) and 1 Narrow Line Seyfert 1 (NLSy1) galaxy. Our spectroscopic campaign will allow us to study the spectroscopic properties (FWHM, EW, flux) of broad-emission lines in the optical (e.g. H$\beta$) and mid-UV (e.g. Mg II $\lambda$2800) regimes, depending on the redshift of the source. The cadence of the broad emission lines monitoring is about five nights per month which in turn will permit us to explore whether there is a correlated variability between broad emission line features and high levels of $\gamma$-ray emission.
\end{small}
\end{abstract}

\maketitle

\thispagestyle{fancy}

\section{INTRODUCTION}
\begin{large}

Among the Active Galactic Nuclei (AGN) blazars are the dominant and brightest population at $\gamma$-rays \citep{Ab:10a}. The $\gamma$-ray emission in blazars is thought to be produced via Inverse Compton (IC) scattering of seed photons from the jet itself (Synchrotron Self Compton or SSC; e.g. \citealp{BM:96a}), or from a source external to the jet (the accretion disk, the dusty torus or maybe the broad line region or BLR; e.g. \citealp{Si:94a}). Many theoretical and observational works have been published over the years, however, we still do not know where does the IC takes place (in the jet, torus, BLR or accretion disk).

Despite all the multiwavelentgth campaigns, just a few efforts in monitoring the spectral variability have been done for a sizeable sample of objects (e.g. The Steward Monitoring Program; \citealp{Steward}). 

The main purpose of this project is to look for evidence that would suggest a response in the BLR to the non-thermal continuum in blazars. Such behavior has been previously found in two radio galaxies (3C 390.3 and 3C 120) with superluminal motions \citep{Ar:10a,LT:10a}. In addition, our team has recently found a flare on the Mg II $\lambda$2800 emission line in 3C 454.3, which is coincident with a superluminal jet component traversing through the radio core, and also coincides with a flare in the non-thermal optical spectral continuum (see Fig.~\ref{mgii}); this finding links the broad emission line fluctuations to the non-thermal continuum emission produced by relativistically moving material in the jet, and hence to the presence of BLR clouds surrounding the radio core \citep{Mio}.

In this project, we aim to explore the variability of the broad emission lines in blazars, in order to use it as an auxiliary piece of information to probe the geometry and physics of the innermost regions of blazars, and to provide evidence for the above scenarios of the $\gamma$-ray production.

\begin{figure}[h]
\includegraphics[width=75mm]{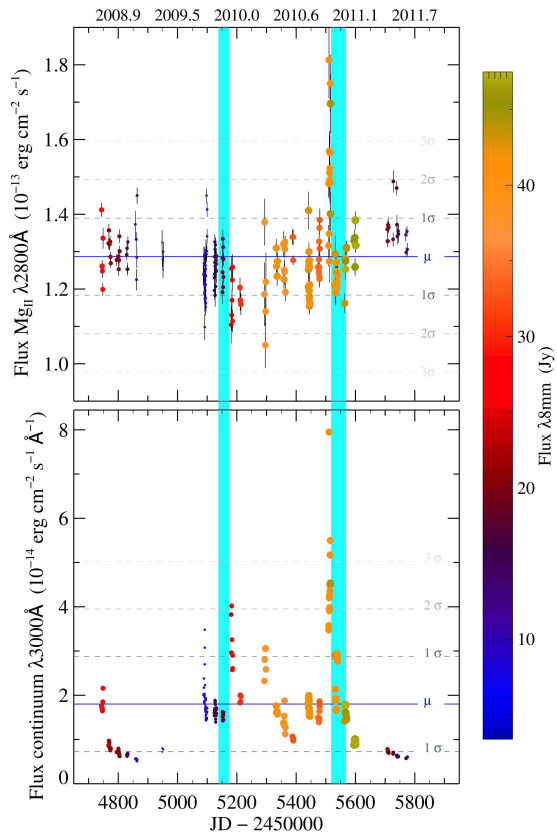}
\caption{Flux evolution of Mg II $\lambda$2800 emission line (top panel) and UV-continuum $\lambda$3000 \AA$\;$emission (bottom panel). For details about these plots see Fig. 3 of \cite{Mio}).}
\label{mgii}
\end{figure}

\section{THE OBSERVATIONAL PROGRAM}

Our spectroscopic monitoring program is designed to homogeneously monitor the broad emission lines in a sample of Fermi bright sources, for about five nights per month. By using a controlled instrumental setup (same telescope, same detector, same slit size, same resolution, etc.) along each observing campaign, we will be able to identify broad-line flux variability (if any) and more importantly, changes in the broad emission line profiles. Studying the broad emission line shape variability will provide important clues on whether the physical conditions of the BLR change during flare episodes. This program will be complemented with observations obtained at different wavelengths.

Our monitoring program is complementary to the monitoring program that is being carried out at the Steward Observatory, University of Arizona \citep{Steward}.

A multiwavelength study of Fermi/LAT blazars is being carried-out (see \citealp{PA:13a}) using data in $\gamma$-rays from Fermi/LAT, near-infrared from the Observatorio Astrof\'isico Guillermo Haro (OAGH, Cananea, Sonora, M\'exico) and SMARTS \citep{SMA:12a}, V band from the Steward Observatory \citep{Steward} and SMARTS as well, millimeter emission and in the near future we will include X-rays. This information will be used to study the connection between the jet, BLR and $\gamma$-ray emission with the aim to construct emission maps, which will allow us to reconstruct the geometry and physics in the central regions of $\gamma$-ray blazars.

We are open to collaborations with other monitoring programs or institutions.

\begin{table}[t]
\begin{center}
\begin{threeparttable}
\caption{The Sample}
\begin{tabular}{lcc}
\hline \textbf{Source} & \textbf{NED Classification} & \textbf{z} \\
\hline \textbf{[HB89] 0235+164} & \textbf{FSRQ} & \textbf{0.940} \\
 \textbf{1H 0323+342} & \textbf{NLSy1} & \textbf{0.061} \\
 \textbf{OJ 248} & \textbf{FSRQ} & \textbf{0.940} \\
 \textbf{[HB89] 0906+015} & \textbf{FSRQ} & \textbf{1.024} \\
 \textbf{[HB89] 1156+295} & \textbf{FSRQ} & \textbf{0.724} \\
 \textbf{3C 273} & \textbf{FSRQ} & \textbf{0.158} \\
 \textbf{3C 279} & \textbf{FSRQ} & \textbf{0.536} \\
 \textbf{[HB89] 1510-089} & \textbf{FSRQ} & \textbf{0.360} \\
 \textbf{S4 1849+67} & \textbf{FSRQ} & \textbf{0.657} \\
 \textbf{[HB89] 2145+067} & \textbf{FSRQ} & \textbf{0.990} \\
 \textbf{3C 454.3} & \textbf{FSRQ} & \textbf{0.859} \\
 \textbf{[HB89] 2255-28} & \textbf{FSRQ} & \textbf{0.923} \\
\hline
\end{tabular}
\label{sample}
\begin{tablenotes}
\item FSRQ: Flat-Spectrum Radio Quasar, NLSy1: Narrow-Line Seyfert 1.
\end{tablenotes}
\end{threeparttable} 
\end{center}
\end{table}

\subsection{Facilities and Instrumentation}

The observations are being carried out using the 2.12 m Telescope at the OAGH; using the Boller \& Chivens Spectrograph. The detector is a 1024$\times$1024 pixel$^2$, SITe CCD having $\lesssim$4 e$^{-}$ read noise\footnote{For more information about the observatory, please refer to \url{http://www.inaoep.mx/~astrofi/cananea/#english}.}. In addition, an extensive Near-Infrared monitoring is being carried out in the same telescope (see \citealp{Ca:09a}).

\begin{figure*}[t]
\includegraphics[width=0.8\textwidth]{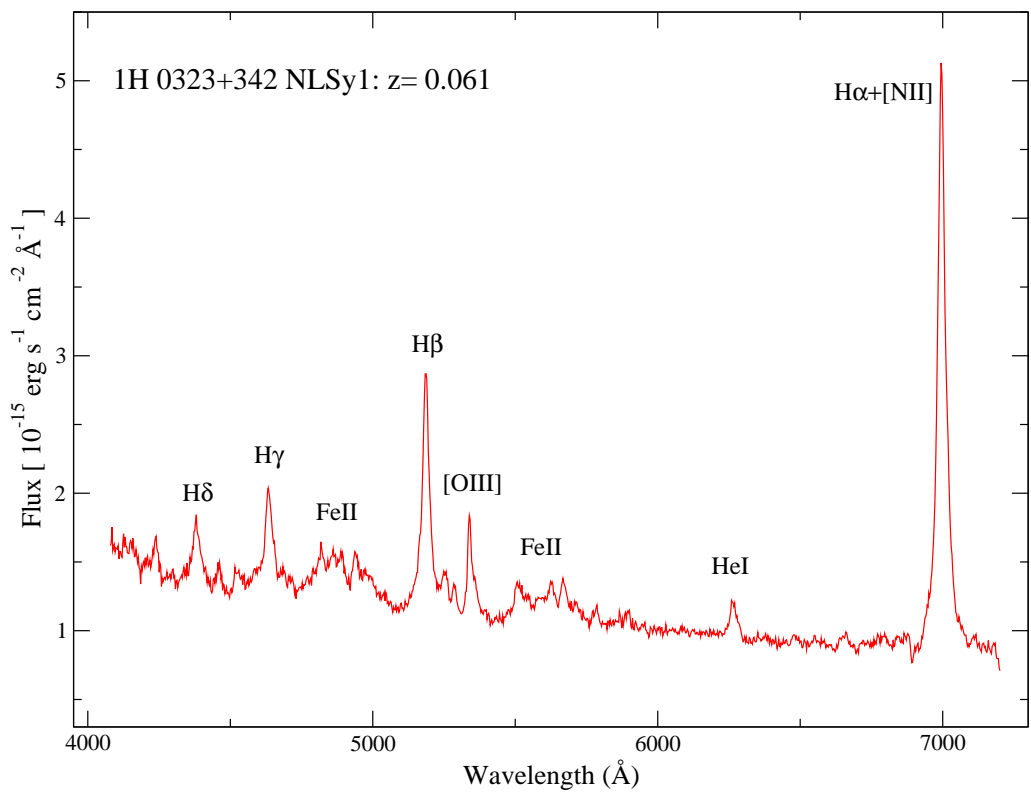}
\caption{Optical spectrum of Narrow Line Sy1 galaxy 1H 0323+342 from October 13th 2012. Observation was carried out at OAGH. The spectral resolution is about 15 \AA.} \label{espectro}
\end{figure*}

\subsection{Data Reduction and Calibration}

The data are being reduced using the IRAF package. We are using comparison spectra of He-Ar to calibrate the spectra in wavelength. For the flux calibration of the spectra we are using spectrophotometric standard stars which we take every night (at least two per night). In the spectra where we have prohibited lines (specially [O III]$\lambda$5007) we can recalibrate the flux of the spectra to obtain a better accuracy (see \citealp{Sh:04a}). In order to obtain information from the broad and narrow lines in the spectra, in most cases, it is necessary to subtract the Fe II emission (optical and UV, depending on the line we want to study) from the spectra, given that the Fe II emission is blended with the broad and narrow lines. The procedure of Fe II subtraction is described in \cite{To:12a}.

\section{STATUS OF THE SPECTROSCOPIC MONITORING CAMPAIGN}

We are obtaining flux calibrated spectra for a sample of objects with low ($\sim$15 \AA$\;$FWHM) and intermediate ($\sim$6 \AA$\;$FWHM) spectral resolution. We have about five nights per month; this cadence of observations provides a good match to the $\gamma$-ray variability data that Fermi can provide for the brightest blazars in the sky. An example of a low resolution spectrum for a NLSy1 galaxy, recently detected by Fermi/LAT is in Figure~\ref{espectro}.


\begin{figure*}[t]
\includegraphics[width=0.8\textwidth]{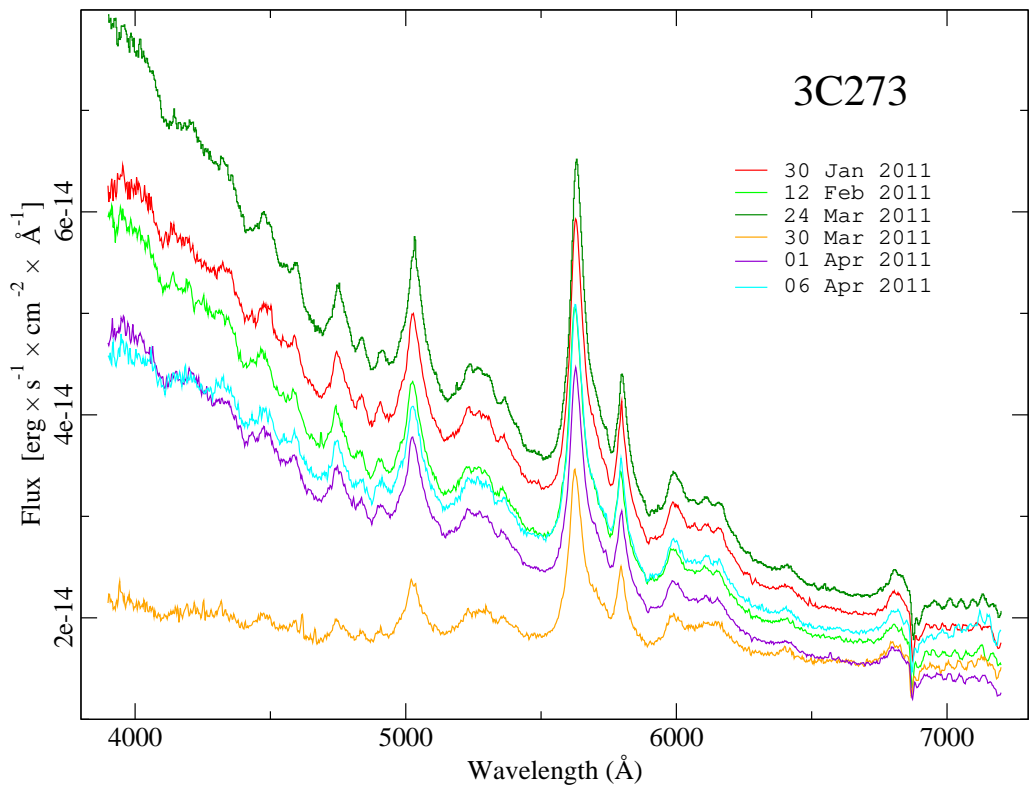}
\caption{Optical spectral variability of the blazar 3C 273 from January to April 2011. Observations were carried out at OAGH. The spectral resolution is about 15 \AA.}
\label{fig_3c273}
\end{figure*}

This program has been running since 2011, and we have an appreciable number of observations for the sources in Table~\ref{sample}. An example of spectral variability we have found can be seen in Fig.~\ref{fig_3c273}.

Besides from the source list, we may also include some other high interest sources depending on the activity phase they are having.


\begin{acknowledgments}

This work was supported by CONACyT research grant 151494 (M\'exico). V.P.-A. acknowledges support from the CONACyT program for PhD studies.

\end{acknowledgments}



\end{large}
\end{document}